\documentclass[namedreferences,natbib]{solarphysics}
\usepackage[optionalrh]{spr-sola-addons} 
\usepackage{graphicx}                    
\usepackage{amssymb}                    
\usepackage[usenames]{color}                       
\usepackage{url}                         

\newcommand{\lb}{\ensuremath{\mathopen{<}}}
\newcommand{\rb}{\ensuremath{\mathopen{>}}}

\usepackage{solaheader}	
\newcommand{\solareceiveddate}{15 Decembre 2011} 
\newcommand{\solaaccepteddate}{6 March 2012}
\begin{document}

\begin{article}

\begin{opening}

\title{The Solar Wind Energy Flux}

%
\author{G.~\surname{Le Chat}$^{1,2}$\sep
        K.~\surname{Issautier}$^{1}$\sep
        N.~\surname{Meyer-Vernet}$^{1}$
       }

%
\runningauthor{G. Le Chat {\it et al.}}
\runningtitle{The Solar Wind Energy Flux}

%
  \institute{$^{1}$ LESIA, Observatoire de Paris, CNRS, UPMC, Universit\'e Paris Diderot; 5 Place Jules Janssen, 92195 Meudon, France
                     email: \url{gaetan.lechat@obspm.fr}\\
                $^{2}$ Harvard-Smithsonian Center for Astrophysics, Cambridge, USA
             }

\begin{abstract}
	The solar-wind energy flux measured near the ecliptic is known to be independent of the solar-wind speed. Using plasma data from {\it Helios}, {\it Ulysses}, and {\it Wind} covering a large range of latitudes and time, we show that the solar-wind energy flux is independent of the solar-wind speed and latitude within $10\%$, and that this quantity varies weakly over the solar cycle. In other words the energy flux appears as a global solar constant. We also show that the very high speed solar wind ($V_\mathrm{SW} >$ 700\,km\,s$^{-1}$) has the same mean energy flux as the slower wind ($V_\mathrm{SW} <$ 700\,km\,s$^{-1}$), but with a different histogram. We use this result to deduce a relation between the solar-wind speed and density, which formalizes the anti-correlation between these quantities.
\end{abstract}

%
\keywords{Plasma Physics; solar wind; Energy Flux}

\end{opening}

%
\section{Introduction}

That the solar wind exists in two basic states, fast and slow, has been known since the first {\it in-situ} observations of the solar wind \citep{Neugebauer62}. Similarities and differences between high- and low-speed solar-wind structures had been extensively studied (see the review of \citet{Schwenn06} and references therein). One of the basic properties of these solar-wind states, together with their differences in composition, variability, and energetics, is the strong anti-correlation between density and flow velocity \citep{Neugebauer66, Hundhausen70, Rosenbauer77, McComas00, Ebert09}. The {\it Ulysses} spacecraft confirmed the existence of the fast solar wind at high latitude during solar minimum, whereas slow solar wind is restricted to near the equatorial plane between about $25^\circ$S and $25^\circ$N \citep{Issautier08}. The fast solar wind has a speed around $750\,\mathrm{km\,s^{-1}}$, a mean density of $2.5\,\mathrm{cm^{-3}}$, an electron temperature of $2\times10^5\,\mathrm{K}$ \citep{LeChat11}, and a proton temperature of $2.3\times 10^5\,\mathrm{K}$ at 1 AU \citep{Ebert09}. For the slow wind at 1 AU, the average speed is around $400\,\mathrm{km\,s^{-1}}$ with a density of $10\,\mathrm{cm^{-3}}$, an electron temperature of $1.3\times 10^5\,\mathrm{K}$ and a proton temperature of $3.4\times 10^4\,\mathrm{K}$ \citep{Schwenn90}. However, despite their large differences in properties and coronal sources, both slow and fast solar wind turn out to have a similar energy flux \citep{Schwenn90, Meyer06, LeChat09}.

In the present article, we use several sets of data from different spacecraft at various heliocentric distances and latitudes to calculate the energy flux during 24 years (Section \ref{data}). We show that the similarity of the mean energy flux between slow and fast wind is a robust property independent of latitude, and that it varies weakly with solar activity and epoch (Section \ref{result}). We also use this property to propose semi-empirical relations between density, velocity and proton temperature in the solar wind and confront them to the data (Section \ref{Discussions}).

\section{Data Analysis}\label{data}

We use the following approximation for the solar-wind energy flux [$W$]:
\begin{equation}\label{W}
	W [\mathrm{W\,m^{-2}}]\,=\,\rho \, V_{\mathrm{SW}} \, \left(\frac{1}{2} V_{\mathrm{SW}}^2\,+\,\frac{M_\odot G}{R_\odot} \right)
\end{equation}
where $\rho$ is the solar-wind density, which is approximated as $\rho=n_\mathrm{p}\,m_\mathrm{p}$ when only protons are considered, or $\rho=n_\mathrm{p}\,m_\mathrm{p}\ +\ n_\mathrm{\alpha}\,m_\mathrm{\alpha}$ when the contribution of ions $\mathrm{He}^{2+}$ ($\alpha$ particles) is taken into account. $V_{\mathrm{SW}}$ is the solar-wind bulk velocity, $M_\odot$ is the solar mass, $R_\odot$ is the solar radius, and $G$ is the gravitational constant. This includes basically the sum of the kinetic energy  of the wind and the energy it needs to leave the Sun's gravitational potential. Equation (\ref{W}) neglects the contribution of the heat flux, of the enthalpy, and of the waves. This is justified by the order of magnitude of these quantities. For instance, \citet{Pilipp90} measured with {\it Helios} an electron heat flux at 1 AU of $q_\mathrm{e}\,\approx\,10^{-6}\ \mathrm{W\,m^{-2}}$. For the protons, \citet{Hellinger11} find $q_\mathrm{p}\,\approx\,10^{-7}\ \mathrm{W\,m^{-2}}$. Compared to our values of $W$ at 1 AU (see Section \ref{result}), both $q_\mathrm{e}$ and $q_\mathrm{p}$ are negligible, as is the contribution of enthalpy and waves \citep{Schwenn90}.

We use one-hour averaged data from {\it Ulysses}/SWOOPS \citep{Bame92} between April 1992 and June 2009, 24-second averaged data from {\it Wind}/ 3DP \citep{Lin95} and one-hour averaged data from {\it Wind}/SWE \citep{Ogilvie95} between November 1994 and September 2011, which allow simultaneous observations at different locations in the heliosphere. The {\it Wind}/3DP and {\it Wind}/SWE data are considered as two independent data sets.Hourly averaged data from the {\it Helios 2}/E1 Plasma Experiment \citep{Rosenbauer77} between January 1976 and March 1980 are also used to provide an earlier temporal comparison point. Since {\it Ulysses} and {\it Helios} orbits exhibit variations in heliocentric distance ($R_{\mathrm{AU}}$), we assume that the solar wind is in spherical expansion at constant speed ($n\,\propto\,R_{\mathrm{AU}}^2$) to scale the density to 1 AU. 

\section{Energy Flux Independence on Latitude and Flow Speed}\label{result}

\subsection{Averaged Values of the Energy Flux}

 \begin{figure} 
 \centerline{\includegraphics[width=1.0\textwidth]{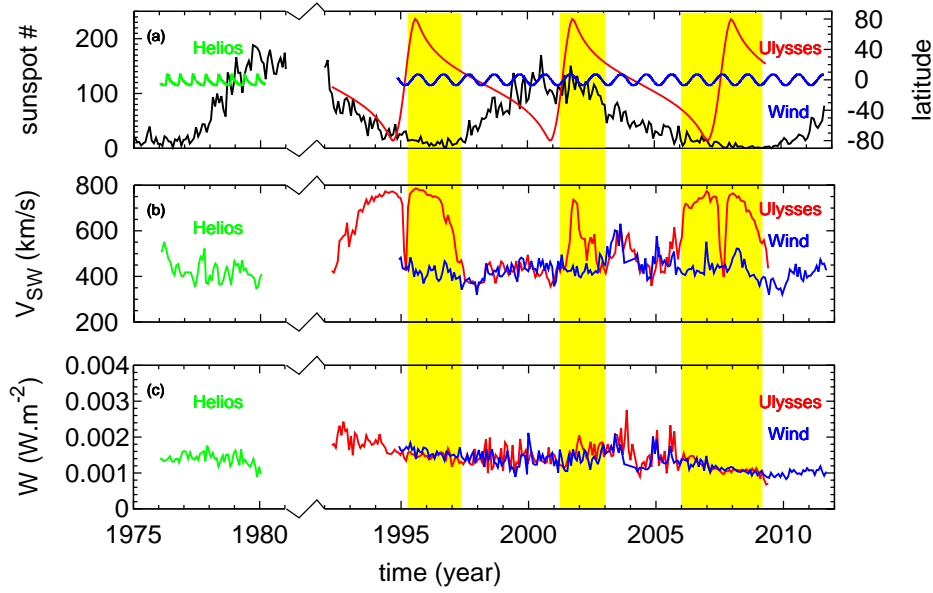}}
 \caption{Panel (a): monthly solar sunspot number superimposed on {\it Helios} (green), {\it Ulysses} (red), and {\it Wind} (blue) heliocentric latitudes. Panel (b): solar-wind speed measured by {\it Helios} (green), {\it Ulysses}/SWOOPS (red) and {\it Wind}/SWE (blue). Panel (c): solar-wind energy flux obtained from Equation (1) for {\it Helios}/E1 Plasma Experiment data (green), {\it Ulysses}/SWOOPS data (red), and {\it Wind}/SWE data (blue). Speed and energy-flux data are averaged over a solar rotation (taken as 27.2 days) and the energy flux is scaled to 1 AU for {\it Helios} and {\it Ulysses}. The time period between the {\it Helios} and {\it Ulysses} epochs have been removed. The yellow bands highlight intervals when {\it Ulysses} and {\it Wind} encounter very different solar-wind conditions and at very different latitudes.}\label{efcompare}
 \end{figure}

Figure \ref{efcompare} (c) shows the energy flux [with $\rho\,=\,n_\mathrm{p}\,m_\mathrm{p}$] obtained from the {\it Helios}/E1 Plasma Experiment, {\it Ulysses}/SWOOPS and {\it Wind}/SWE data. We compare it to the solar-wind speed measurements (Figure \ref{efcompare} (b)), the solar activity represented by the monthly sunspot number \citep{sidc} and the latitude of each spacecraft (Figure \ref{efcompare} (a)). The  energy flux has been calculated from Equation (1) using hourly averaged data, and then averaged over a solar rotation (taken as 27.2 days) to reduce the effect of transient events such as CMEs or CIRs. The averaged energy flux measured by the three spacecraft is $(1.5 \pm 0.4) \times 10^{-3}\,\mathrm{W\,m^{-2}}$ at 1 AU, compatible with the value previously found by \citet{Schwenn90}. The mean values at 1 AU for each spacecraft are respectively: $(1.4 \pm 0.2) \times 10^{-3}\,\mathrm{W\,m^{-2}}$ for {\it Helios}, $(1.7 \pm 0.4) \times 10^{-3}\,\mathrm{W\,m^{-2}}$ for {\it Ulysses}, and $(1.3 \pm 0.3) \times 10^{-3}\,\mathrm{W\,m^{-2}}$ for {\it Wind}. Thus, the energy flux measured by {\it Helios} is compatible with those measured decades later by {\it Ulysses} and {\it Wind}. 

A very remarkable result shown in Figure \ref{efcompare} is that the solar-wind energy fluxes measured by {\it Ulysses} and {\it Wind} follow the same variations during overlapping time periods, almost 16 years, and have similar mean values: $(1.5 \pm 0.4) \times 10^{-3}\,\mathrm{W\,m^{-2}}$ for {\it Ulysses}, and $(1.4 \pm 0.3) \times 10^{-3}\,\mathrm{W\,m^{-2}}$ for {\it Wind}. This leads to a difference of less than $10\%$ despite the different trajectories of the two spacecraft. This is especially interesting in the time periods when {\it Ulysses} and {\it Wind} are in very different solar-wind states and at different latitudes (indicated with yellow zones on Figure \ref{efcompare}). Indeed, this implies that the fast and slow solar winds have the same mean energy flux, either in solar activity maximum (in 2001) or minimum (in 1996 or 2008), and that this invariance is a global solar property, independent of heliolatitude. Furthermore, the highest differences between {\it Ulysses} and {\it Wind} values occur in 2004, when both spacecraft were at similar latitudes (but not at the same distance from the Sun).

Figure \ref{efcompare} shows a long-term variation in energy flux, with a maximum value $50\%$ larger than the minimum value and a periodicity of about 11 years, with a time shift of about three years compared to the solar activity cycle. However, since the amplitude of this long-term variation is of the same order of magnitude as variations between two consecutive solar rotations, we need longer (of at least another solar cycle) continuous survey of the solar-wind energy flux to confirm the 11-year period of this long-term variation.  

\subsection{Energy Flux in Minimum and Maximum of Solar Activity}\label{histoS}

 \begin{figure} 
 \centerline{\includegraphics[width=1.0\textwidth]{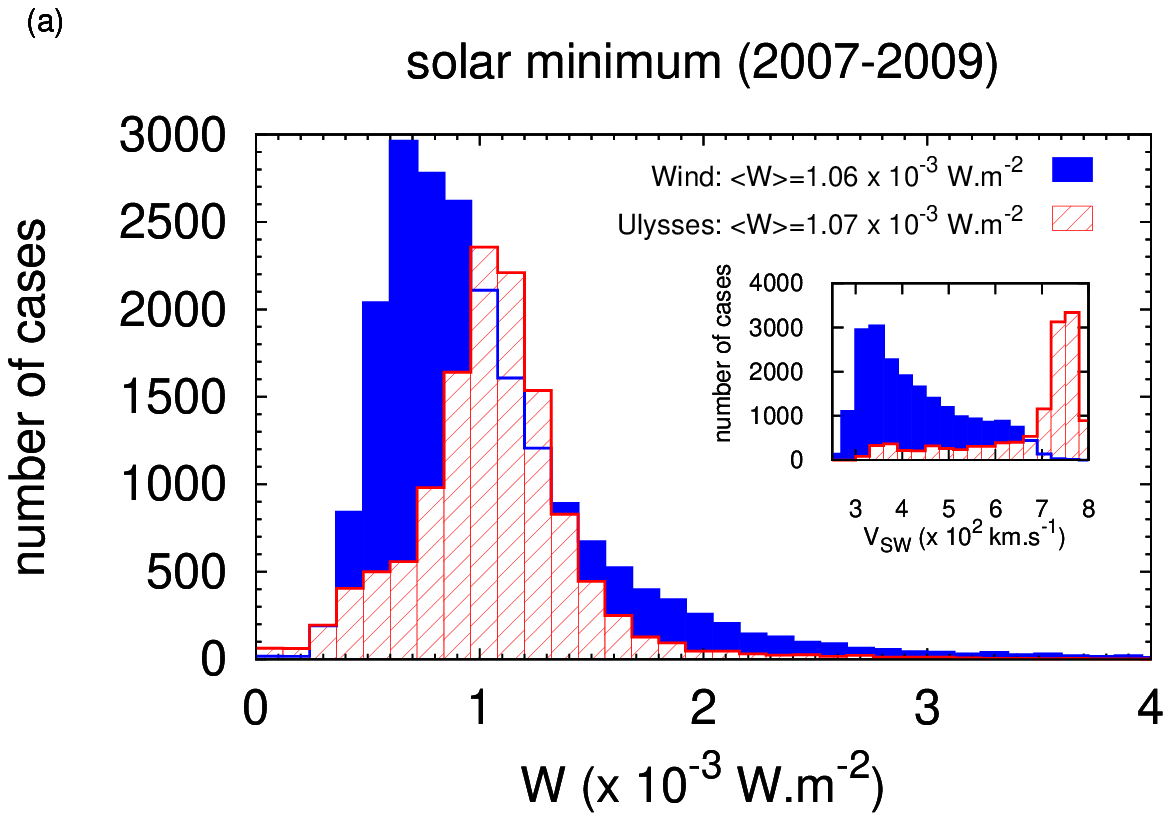}} \centerline{\includegraphics[width=1.0\textwidth]{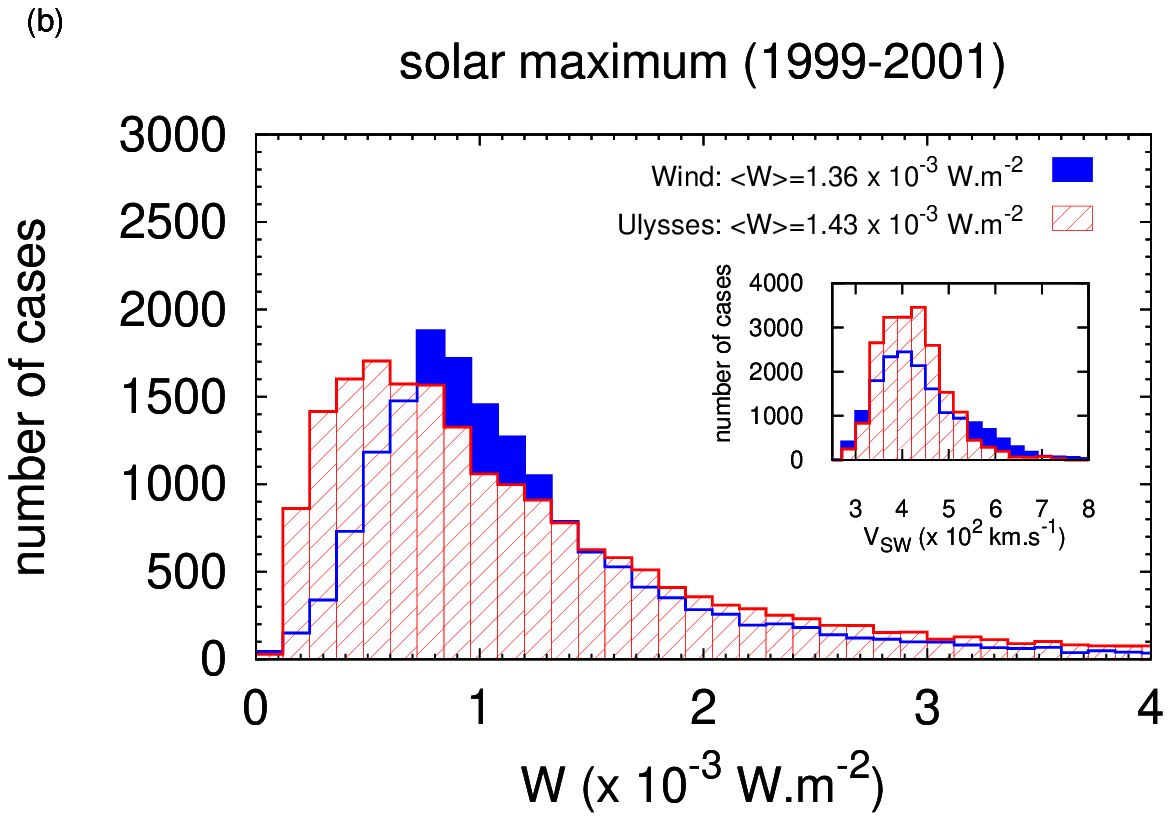}}
 \caption{Histograms of the energy flux measured by {\it Ulysses}/SWOOPS (hatched red) and {\it Wind}/SWE (plain blue) during minimum (panel (a)) and maximum (panel (b)) of solar activity. The energy flux is scaled to 1 AU for {\it Ulysses}. Average values of the energy flux and histograms of the solar-wind speed are given for the time periods considered.}\label{efhisto}
 \end{figure}

Figure \ref{efhisto} compares the histograms of the one-hour averaged energy flux measured by {\it Ulysses}/ SWOOPS (red) and {\it Wind}/SWE (blue) during minimum (panel (a)) and maximum (panel (b)) of solar activity.

The minimum period corresponds to the years 2007 to 2009 when {\it Ulysses} made its last fast pole-to-pole scan of the heliosphere. One can see on Figure \ref{efcompare} that during this period, {\it Ulysses} and {\it Wind} encountered very different solar wind states except during the crossing of the Ecliptic by {\it Ulysses}. Indeed, {\it Ulysses} measured high-latitude fast-wind, with a mean velocity of $678\,\mathrm{km\,s^{-1}}$, while {\it Wind} measured both slow and fast wind, with a dominance of slow solar wind, with a mean velocity of $430\,\mathrm{km\,s^{-1}}$ (see inner histogram in panel (a)). The corresponding histograms of the one-hour averaged energy flux are very dissimilar. In the case of {\it Ulysses}, the distribution is narrow and nearly symmetrical with a most probable value at $1.05 \times 10^{-3}\,\mathrm{W\,m^{-2}}$, corresponding to the energy flux of the high latitude fast solar wind (at a speed of $750 \pm 50  \,\mathrm{km\,s^{-1}}$). On the other hand, the distribution of the energy flux measured by {\it Wind} is very asymmetric, with a peak around $0.7 \times 10^{-3}\,\mathrm{W\,m^{-2}}$. This leads to a difference between the two peaks of more than $30\%$. Nevertheless, the average values during this period are respectively $1.06 \times 10^{-3}\,\mathrm{W\,m^{-2}}$ for {\it Wind} and $1.07 \times 10^{-3}\,\mathrm{W\,m^{-2}}$ for {\it Ulysses}, leading to a difference of less than $1\%$. It is noteworthy that a similar result is found during the 1996 solar-activity minimum, but with a higher value of the energy flux.



The maximum period corresponds to the years 1999 to 2001, when both {\it Ulysses} and {\it Wind} measured slow and fast wind, with a dominance of slow solar wind (inner histogram of Figure \ref{efhisto} (b)), but at different latitudes and distances from the Sun. The distributions of the energy flux are asymmetric for both {\it Ulysses} and {\it Wind}, with  most probable values of $0.55  \times 10^{-3}\,\mathrm{W\,m^{-2}}$ and $0.75  \times 10^{-3}\,\mathrm{W\,m^{-2}}$ for {\it Ulysses} and {\it Wind}, respectively. The average values of the energy flux during this period are respectively $1.36 \times 10^{-3}\,\mathrm{W\,m^{-2}}$ for {\it Wind} and $1.43 \times 10^{-3}\,\mathrm{W\,m^{-2}}$ for {\it Ulysses}, which amounts to a difference of $5\%$, much smaller than the difference in the most probable values.

Co-rotating Interaction Regions (CIRs) appear to be the main cause of the differences between the distributions of the energy flux measured by {\it Ulysses} and {\it Wind}. Indeed, the high-latitude fast solar wind seen by {\it Ulysses} during solar minimum does not interact with slower streams. This can explain the nearly Gaussian distribution shown in Figure \ref{efhisto} (a) for {\it Ulysses}, compared to the asymmetric distribution observed by {\it Wind} in the Ecliptic (where fast and slow winds interact). Note that only few CMEs were observed by {\it Wind} during this period (representing only eight hours of this two-year data set, namely 0.5\% of the time corresponds to CMEs plasma). CIRs can also explain the differences found in the distributions of the energy flux during maximum of solar activity. Since CIRs are stronger the farthest away from the Sun they are observed, {\it Ulysses} measurements are more affected by CIRs than {\it Wind} data.

\subsection{Contribution of $\alpha$ Particles to the Energy Flux}

At this point of this article and in the previous study by \citet{Schwenn90}, only protons are considered in Equation (\ref{W}) [$\rho=n_\mathrm{p}\,m_\mathrm{p}$]. During the {\it Ulysses} and {\it Wind} overlapping time periods, on average, the $\alpha$ particles increase the energy flux by $15\%$, but they do not change the similarity between fast and slow solar wind. The averaged value of the energy flux that we obtained is $\lb W \rb\,=\,(1.7\pm 0.4)\times 10^{-3}\,\mathrm{W\,m^{-2}}$ at 1 AU. Assuming a scaling as $R_\mathrm{AU}^2$, this would correspond to an energy flux at one solar radius of $79\,\mathrm{W\,m^{-2}}$ as previously found by \citet{LeChat09}. In the next sections, we will use $\rho=n_\mathrm{p}\,m_\mathrm{p}\ +\ n_\mathrm{\alpha}\,m_\mathrm{\alpha}$.

\section{Semi-empirical Relation between Speed and Density}\label{Discussions}


We have shown in Section \ref{result} that on average the solar-wind energy flux is independent of the solar-wind velocity and heliolatitude. It is straightforward to derive a relation between the solar-wind mass density and the solar-wind speed from Equation (\ref{W}):

\begin{equation}\label{relation}
	\rho\ =\ W \left[ V_{\mathrm{SW}}\left(\frac{V_{\mathrm{SW}}^2}{2} + \frac{M_\odot G}{R_\odot}\right) \right]^{-1}
\end{equation}

For the {\it Wind}/3DP 24-second averaged data set corresponding to more than 1.9 billion measurements from December 1994 to September 2011, the following most probable value of the solar energy flux at 1 AU is: $$\widetilde{W}\,=\,8.5 \times 10^{-4}\mathrm{W\,m^{-2}}$$ with the corresponding full width at half maximum: $$[\sigma_-:\sigma_+]\,=\, [4.1 : 16] \times 10^{-4}\,\mathrm{W\,m^{-2}}$$ where $\sigma_-$ and $\sigma_+$ designate the lower and upper boundaries of the full width at half maximum respectively. Using this value of $\widetilde{W}$, Equation (\ref{relation}) becomes:

\begin{equation}\label{prelation}
	\rho\ \approx\ 1.7\times 10^{-12} \left[ V_{\mathrm{SW}} (V_{\mathrm{SW}}^2 + 3.81\times 10^5) \right]^{-1}
\end{equation}

\noindent with $\rho=n_\mathrm{p}\,m_\mathrm{p}\ +\ n_\mathrm{\alpha}\,m_\mathrm{\alpha}$ in $\mathrm{kg\,m^{-3}}$ and $V_{\mathrm{SW}}$ in $\mathrm{km\,s^{-1}}$. The values of $\sigma_-$ and $\sigma_+$ allow to estimate the following confidence interval of the first numerical value of Equation (\ref{prelation}): $[8.2 \times 10^{-13} : 3.2 \times 10^{-12}]$. 

 \begin{figure} 
 \centerline{\includegraphics[width=1.0\textwidth]{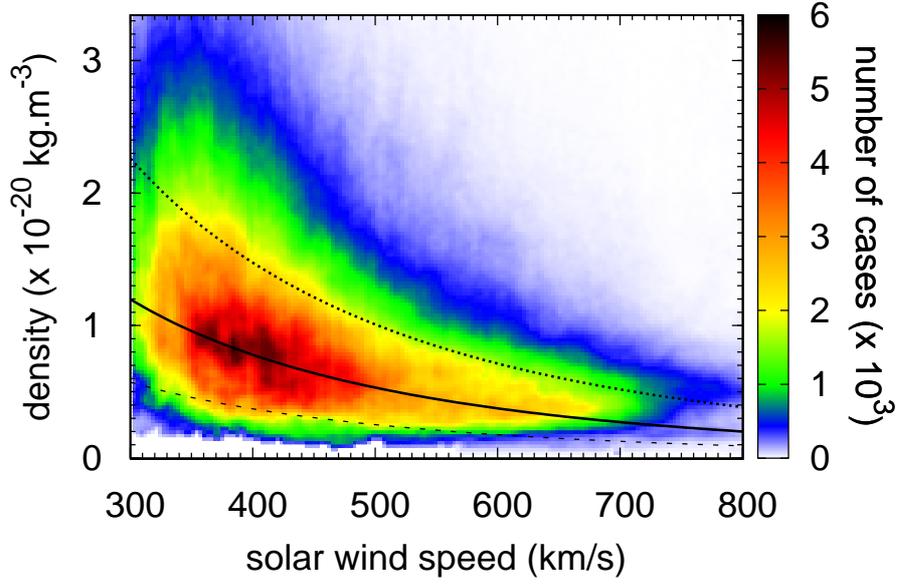}}
 \caption{Density {\it versus} the solar-wind speed measured by {\it Wind}/3DP, with the corresponding color bar chart on the right side of the figure. This data set corresponds to more than 1.9 billion measurements from December 1994 to September 2011. The solid line is the theoretical relation obtained from the most probable value of the energy flux.The dashed and dotted lines are the relations obtained using respectively the lower and upper  boundaries of the full width at half maximum of $W$.}\label{nVrelation}
 \end{figure}

Figure \ref{nVrelation} compares the solar-wind density and speed measured at 1 AU to Equation (\ref{prelation}). This shows that this relation represents very well the data up to $700\,\mathrm{km\,s^{-1}}$. Between 700 and 800 $\mathrm{km\,s^{-1}}$, the distribution of density starts to differ with the prediction of Equation (3). A similar figure can be obtained for each data sets we used in this paper after scaling to 1 AU the {\it Helios} and {\it Ulysses} data.

Equation (\ref{prelation}) can be coupled with the well-known statistical correlation between solar-wind speed [$V_\mathrm{SW}$] and proton temperature [$T_\mathrm{p}$] in order to deduce an empirical relation between the solar-wind density and the proton temperature. One of the earliest study of the $V_\mathrm{SW}-T_\mathrm{p}$ relation was done by \citet{Hundhausen70}, who concluded that {\it Vela 3} data could be fitted with a linear fit of either $\sqrt{T_\mathrm{p}}$ or $T_\mathrm{p}$ as a function of $V_\mathrm{SW}$. Since then, both linear and quadratic fits have been used \citep{Burlaga70, Lopez86, Lopez87, Richardson95, Neugebauer97, Neugebauer03, Elliot05}. We choose to use the relation obtained by solving the internal energy and momentum equations (Equation (23) of \citet{Demoulin09}). It implies, for a given distance and heating flux, that the proton temperature is a quadratic function of the velocity. At 1 AU, comparison to the data of \citet{Matthaeus06} leads to the following relation: $T_\mathrm{p}\,\approx\,0.5 V_\mathrm{SW}^2$, with $T_\mathrm{p}$ in K and $V_\mathrm{SW}$ in $\mathrm{km\,s^{-1}}$. Consequently, the relation between $\rho$ and $T_\mathrm{p}$ is:

\begin{equation}\label{trelation}
	\rho\,\approx\, 6\times 10^{-13} \left[ \sqrt{T_{\mathrm{p}}} (T_{\mathrm{p}} + 1.9\times 10^5) \right]^{-1}
\end{equation}

\noindent with $\rho$ in $\mathrm{kg\,m^{-3}}$, and $T_\mathrm{p}$ in K.

Table (\ref{sum}) gives some typical values of the solar-wind properties using Equations (\ref{prelation}) and (\ref{trelation}). The values obtained are compatible with those previously published (\citet{Ebert09}, and others).

\begin{table}
 \caption{{Most probable values and confidence intervals of $\rho$, $n$, and $T_p$ for both slow ($V_\mathrm{SW}=400\,\mathrm{km\,s^{-1}}$) and fast ($V_\mathrm{SW}=750\,\mathrm{km\,s^{-1}}$) solar wind obtained using Equations (\ref{prelation}) and (\ref{trelation}).}}\label{sum}
 \begin{tabular}{l c c}     
 	\hline\noalign{\smallskip}
	& slow wind & fast wind\\
 	\hline
	$V_\mathrm{SW}$ [$\mathrm{km\,s^{-1}}$] & 400 & 750\\
	$\rho$ [$\times 10^{-20}\,\mathrm{kg\,m^{-3}}$] & $0.8\ (\in [0.4\,:\,1.5])$ & $0.24\ (\in [0.12\,:\,0.45])$\\
	$n$ [$\mathrm{cm^{-3}}$] & $4.7\ (\in [3.8\,:\,8.8])$ & $1.4\ (\in [1.0\,:\,4.5])$\\
	{$T_\mathrm{p}$ [$\times 10^5$ K]} & {$0.8\ (\in [0.3\,:\,1.7])$} &{ $2.8\ (\in [1.5\,:\,5.1])$}\\
	\hline\noalign{\smallskip}
 \end{tabular}
 \end{table}


\section{Discussions and Conclusions}

\begin{figure} 
 \centerline{\includegraphics[width=0.5\textwidth]{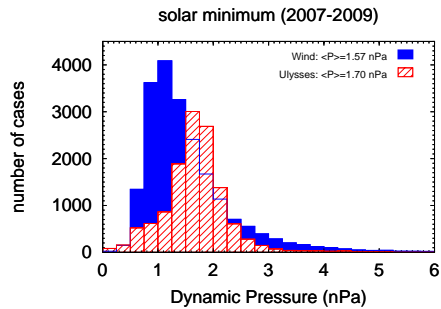}}
 \caption{Comparison between the histograms of the dynamic pressure measured by {\it Ulysses} (hatched red) and {\it Wind} (plain blue) during minimum of solar activity. Averaged values of the dynamic pressure are given for the time periods considered. The histogram of the solar-wind speed is the same as the inner histogram of panel (a) Figure \ref{efhisto}}\label{dynhisto}
 \end{figure}

Using 24 years of solar-wind data from {\it Helios}, {\it Ulysses}, and {\it Wind}, we find that the average solar-wind energy flux is independent of heliolatitude and similar for both fast and slow solar wind. Furthermore, this quantity varies weakly over the solar cycle, so that the solar-wind energy flux appears as a global solar constant. This result generalizes a previous finding based on data sets restricted to low latitudes \citep{Schwenn90}. Nevertheless, it is interesting to note that even if the mean value is very similar, the distribution of the energy flux for the very high speed solar wind ($V_{SW}=750\pm50\,\mathrm{km\,s^{-1}}$) is different than the one of the solar wind at speed below $700\,\mathrm{km\,s^{-1}}$. This very high speed solar wind is mostly observed by {\it Ulysses} at high latitudes during solar-activity minimum and corresponds to a steady-state solar wind without interactions with slower streams. Figure \ref{dynhisto} shows that a similar result holds for the dynamic pressure, whose global invariance has been previously observed \citep{Steinitz83, Schwenn90, richardson1999}. The mean value of the dynamic pressure is similar for {\it Ulysses} and {\it Wind} experiencing very different solar-wind speeds and latitudes, but the distributions are different.

Given the different sources of the slow and fast wind, and the large difference generally assumed in the respective expansion factors of their flux tubes, the global nature of the invariance of the energy flux is very puzzling. It appears as if the energy flux provided all over the surface of the Sun, the flux-tube expansion, and the interaction between different streams work together in order to yield the same energy flux at large distances.

In this article, we give a direct relation between the solar-wind speed and its density using the invariance of the energy flux. This relation agrees with almost 17 years of continuous {\it in-situ} measurements. Consequently, this relation, which formalizes the well-known anti-correlation between the solar-wind speed and density, can be used as a proxy for deducing the speed from the density.

\citet{LeChat09} had also shown that a large spread of stellar winds, including solar-like and cool-giant stars, have a similar value for their stellar-wind energy flux, suggesting that a shared fundamental process might be at the origin of stellar winds.

%

%


%

%
 \begin{acks}
 The authors thank ESA and the SWOOPS instrument team (D. McComas, PI) for making {\it Ulysses} plasma data available on the ESA web site, and W. Olgivie, A.J. Lazarus, and M.R. Aellig for the SWE data. The {\it Wind}/3DP and {\it Helios} data were provided by the CDAWeb managed by NASA.
 \end{acks}

%
%
 \bibliographystyle{spr-mp-sola}
 \bibliography{SOLA1815R3_Le_Chat}  
%
%
%
%

\end{article} 
\end{document}